\documentclass[epj]{svjour}
\usepackage{graphics}
\begin{document}
\title{Chiral extrapolation of lattice data for $B$-meson decay constant}

\author{X.-H. Guo \thanks{email: xhguo@bnu.edu.cn}\and M.-H. Weng
}                     
\institute{Institute of Low Energy Nuclear Physics, Beijing Normal
University, Beijing 100875, China}
\date{Received: date / Revised version: date}

\abstract{ The $B$-meson decay constant $f_B$ has been calculated
from unquenched lattice QCD in the unphysical region. For
extrapolating the lattice data to the physical region, we propose
a phenomenological functional form based on the effective chiral
perturbation theory for heavy mesons, which respects both the
heavy quark symmetry and the chiral symmetry, and the
non-relativistic constituent quark model which is valid at large
pion masses. The inclusion of pion loop corrections leads to
nonanalytic contributions to $f_B$ when the pion mass is small.
The finite-range regularization technique is employed for the
resummation of higher order terms of the chiral expansion. We also
take into account the finite volume effects in lattice
simulations. The dependence on the parameters and other
uncertainties in our model are discussed.
} 
\maketitle
\section{Introduction}
\label{intro} An accurate determination of the Cabibbo - Kobayashi
- Maskawa (CKM) matrix of the Standard Model and tests of its
consistency and unitarity constitute an important part of current
research in particle physics. Among the most important matrix
elements is the $B$-meson decay constant $f_B$, which is needed to
determine the CKM matrix elements such as $V_{td}$. Lattice
Quantum Chromodynamics (QCD) simulations, which provide a way to
determine $f_B$ from the first principles of QCD, are of
fundamental importance since the determination of the $B$-meson
decay constant still remains beyond the reach of experiment.

In Ref.\cite{Alan}, the authors studied $f_B$ extensively from
full and partially quenched lattice QCD. In the simulations they
applied two actions, the unquenched gauge configurations by the
MILC Collaboration\cite{C.B} and the improved staggered light
quark action\cite{S.N}, for simulating the light quarks ($u, d,
s$). The non-relativistic QCD (NRQCD) formalism was used to treat
the heavy quarks ($c, b$)\cite{B.A}. The lattice data were
obtained in the region where the mass of light quark is about in
the region 20 MeV $\sim$ 80 MeV\cite{Alan} (corresponding to the
pion mass between around 250 MeV and 500 MeV), which is larger
than its physical value. The light quark masses in these
simulations are much closer to the physical region than previous
work. This makes it possible to reduce the uncertainty from chiral
extrapolation. To extrapolate the lattice data to the physical
region the staggered chiral perturbation theory was employed in
Refs.\cite{Alan}\cite{C.A}.

In QCD, when the heavy quark mass goes to infinity, the strong
interaction can be described by the heavy quark effective theory
(HQET)\cite{Mark98}, which contains the heavy quark symmetry
$SU(2)_f \times SU(2)_s$. On the contrary, when the masses of
light quarks $(u,d,s)$ approach zero, the strong interaction has
$SU(3)_L \times SU(3)_R$ chiral symmetry which is spontaneously
broken to the $SU(3)_V$ subgroup, leading to eight pseudoscalar
Goldstone bosons. When the light quark mass (or the pion mass) is
small enough, the effective chiral Lagrangian\cite{Mark93}
containing the chiral and heavy quark symmetries may be applied,
based on power counting which is the foundation of chiral
perturbation theory. The authors of Ref.\cite{thomas} found that
the regime where power counting may be applied is $0 \leq m_\pi
\leq 180$ MeV by analyzing the chiral extrapolation of the nucleon
mass to fourth-order in the expansion and demanding the accuracy
of the prediction for the nucleon mass to be one percent. This
regime is beyond the reach of current lattice simulations. In
order to extrapolate lattice data which are outside the
power-counting regime, the so-called finite-range regularization
(FRR) technique is proposed\cite{thomas,adelaide,guo1,guo2}, in
which a finite-range cutoff (which is physically of order of the
inverse size of the pion source) in the momentum integrals of the
pion loop diagrams is introduced for the resummation of the chiral
expansion. It has been shown strictly that FRR is mathematically
equivalent to minimal subtraction schemes such as dimensional
regularization to any finite order\cite{thomas}. Furthermore, the
model dependence associated with the shape of the regulator is at
one percent level in the range $0 \leq m_\pi \leq 1$ GeV, which is
well outside the power-counting regime. In the present work we
will adopt FRR while applying the effective chiral Lagrangian to
extrapolate the lattice data for the $B$-meson decay constant
since these data are outside the power-counting regime.

The constituent quark model has been shown to work quite well
although it is a very simple model where Quantum Chromodynamics is
not employed. At large pion masses, we employ the non-relativistic
constituent quark model to fit the lattice data. In fact, for all
the hadron properties (such as heavy meson masses and decay
constants) which have been calculated in lattice QCD, the lattice
data vary slowly and smoothly when the light quark mass is larger
than 60 MeV or so. This is characteristic of constituent quark
behavior and suggests that in this region the constituent quark
model should be most appropriate\cite{cloet}. In other words, the
hadron properties vary as a function of the constituent quark mass
in the region of heavier quark mass. It has been shown that the
constituent quark model is consistent with the modeling of QCD via
the Dyson-Schwinger Equation in ladder-rainbow truncation when the
quark mass is large\cite{guo2}. Obviously the lattice data at
large light quark masses in Ref.\cite{Alan}  appear in the region
where the constituent quark model is suitable to be employed.
Consequently, we apply the constituent quark model  in this large
quark mass region and then analytically continue the hadron
properties to the physical mass regime in a manner consistent with
chiral symmetry, because of which nonanalytic terms are involved.

In Ref.\cite{Alan}, the authors extrapolated the lattice data for
the quantity $\Phi_B$ ($\Phi_B = f_B \sqrt{m_B}$) with the
following formula:
\begin{equation}
\Phi_B = c_0 (1 + \Delta + {\rm analytic \;\; terms}),
\label{alan}
\end{equation}
where $c_0$ is a constant associated with the axial-vector current
which destroys $B$ meson, $\Delta$ is the term encompassing the
chiral logarithms, and the analytic term is linear in the light
quark mass. It is obvious that when Eq.(\ref{alan}) is used for
extrapolating the lattice data, power counting in chiral
perturbation theory has been applied when the light quark mass (or
the pion mass) is small and the term linear in the light quark
mass has been assumed when the light quark mass is large. However,
as mentioned before, the lattice data appear in the light quark
mass range 20 MeV $\sim$ 80 MeV  (or the pion mass range 250 MeV
$\sim$ 500 MeV). This is already outside the regime where power
counting can be applied ($0 \leq m_\pi \leq 180$) for small pion
masses and inside the regime where the constituent quark model is
most appropriate for large pion masses. Therefore, based on our
above argument, instead of using the form like Eq.(\ref{alan}) to
extrapolate the lattice data for the $B$-meson decay constant, we
adopt FRR when applying chiral Lagrangian and the constituent
quark model at small and large pion masses, respectively.

The outline of this paper is as follows. We will give a brief
review for the chiral perturbation theory for heavy mesons in
Sect. 2. In Sect. 3, we will propose a functional form for
extrapolating the lattice data to the physical region based on the
effective chiral perturbation theory for heavy mesons and the
constraints from the constituent quark model at large pion masses.
Then in Sect. 4, we will present our fits to the lattice data and
give the numerical results. Finally, a summary and discussion will
be given in Sect. 5.

\section{Chiral perturbation theory for heavy mesons}
\label{sec:1} The chiral perturbation theory for heavy mesons,
which describes the interactions of mesons containing a single
heavy quark with light pseudoscalar bosons, contains both the
heavy quark symmetry $SU(2)_f \times SU(2)_s$ and the chiral
symmetry $SU(3)_L \times SU(3)_R$, provided $m_Q\rightarrow
\infty$ and $m_q\rightarrow 0$ ($Q$ = $c$ or $b$, $q = u, d, s$ ),
respectively\cite{Mark93}.

The pseudoscalar Goldstone bosons are incorporated in a $3 \times
3$
 unitary matrix
\begin{equation}
\Sigma = {\rm exp} \left( \frac{2iM}{f_\pi} \right), \label{1a}
\end{equation}
where $f_\pi$ is the pion decay constant, $f_\pi\simeq 131$ MeV,
and
\begin{eqnarray}
M=\left (
\begin{array}{ccc}
\frac{1}{\sqrt{2}}\pi^0+\frac{1}{\sqrt{6}}\eta & \pi^+ & K^+ \\
\pi^- &-\frac{1}{\sqrt{2}}\pi^0+\frac{1}{\sqrt{6}}\eta &K^0\\
K^- &\bar{K}^0 &-\sqrt{\frac{2}{3}}\eta\\
\end{array}
\right). \label{2b}
\end{eqnarray}

While discussing the strong interaction between heavy mesons and
light pseudoscalar mesons it is convenient to introduce
\begin{equation}
  \xi = \sqrt{\Sigma}.
  \label{xi}
\end{equation}

Under a chiral $SU(3)_L \times SU(3)_R$ transformation,
\begin{eqnarray}
  \Sigma &\rightarrow & L\Sigma R^\dag, \nonumber\\
\xi &\rightarrow & L \xi U^\dag = U\xi R^\dag,\nonumber
\end{eqnarray}
where the unitary matrix $U$ is a complicated nonlinear function
of $L$, $R$ and the pseudoscalar Goldstone boson fields.

In order to describe the interactions of the Goldstone bosons with
the heavy mesons containing $Q\bar{q}^a$ (here $a = 1,2,3$ for
$u,d,s$ quarks, respectively), it is convenient to introduce a
$4\times 4$ matrix  $H_a$ given in Ref.\cite{Mark93},
\begin{equation}
H_a = \frac{1+\rlap/v}{2}(P_{a\mu}^*\gamma^\mu-P_a \gamma_5),
\label{5}
\end{equation}
where $P_a$ and $P^*_{a\mu}$ are the field operators that destroy
a heavy pseudoscalar meson ($P_a$) and a heavy vector meson
($P^*_{a}$)  with four-velocity $v$, respectively. $P_{a\mu}^*$
satisfies the constraint as the following:
\begin{equation}
v^\mu P_{a\mu}^* = 0 \label{6}.
\end{equation}

Under $SU(3)_L \times SU(3)_R$ transformation,
\begin{equation}
  H_a \rightarrow H_b U_{ba}^\dag,\nonumber
\end{equation}
and under the heavy quark spin transformation,
\begin{equation}
  H_a \rightarrow SH_a.\nonumber
\end{equation}

Defining
\begin{equation}
\bar{H}_a =\gamma^0 H_a^\dag \gamma^0,
\end{equation}
we have
\begin{equation}
\bar{H}_a =
(P_{a\mu}^\dag\gamma^\mu+P_a^\dag\gamma_5)\frac{1+\rlap/v}{2}.
\label{10}
\end{equation}

It is convenient to introduce a vector field $V_{ab}^\mu$,
\begin{equation}
 V_{ab}^\mu =
\frac{1}{2}(\xi^\dag\partial^\mu\xi +
\xi\partial^\mu\xi^\dag)_{ab},\label{vmu}
\end{equation}
and an axial-vector field $A_{ab}^\mu$,
\begin{equation}
A_{ab}^\mu = \frac{i}{2}(\xi^\dag\partial^\mu\xi -
\xi\partial^\mu\xi^\dag)_{ab},\label{a}
\end{equation}
to establish the Lagrangian for the chiral perturbation theory.

The most general form of the effective Lagrangian density, which
is invariant under the heavy quark symmetry and the chiral
symmetry and should be invariant under Lorentz and parity
transformations as well, is as follows\cite{Mark93}:
\begin{equation}
{\cal L}= -{\rm Tr}[\bar{H}_a iv_\mu (D^\mu H)_a]+g {\rm
Tr}(\bar{H}_a H_b \gamma_\mu A^\mu_{ba} \gamma_5), \label{l1}
\end{equation}
where $g$ is the coupling constant describing the interaction
between heavy mesons and Goldstone bosons, which contains
information about the interaction at the quark level. So it cannot
be fixed from the chiral perturbation theory for heavy mesons, but
should be determined by experiments. The covariant derivative in
Eq.(13) is defined as
\begin{equation}
(D^\mu H)_a =
\partial^\mu H_a - H_b V_{ba}^\mu.
\end{equation}

Then we have the following explicit form for the interaction of
heavy mesons with Goldstone bosons after substituting
Eqs.(\ref{5})(\ref{10})(\ref{vmu})(\ref{a}) into Eq.(\ref{l1}):
\begin{eqnarray}
 &{\rm Tr}[\bar{H}_a iv_\mu V^\mu_{ba}
H_b]+g {\rm Tr}(\bar{H}_a H_b
\gamma_\mu A^\mu_{ba} \gamma_5)\nonumber \\
=& \frac{i}{f_\pi^2}v^\mu
[M,\partial_\mu M]_{ba}(P_{a \nu}^{* \dag} P_{b}^{* \nu}- P_a^\dag P_b) \nonumber\\
& -\frac{2g}{f_\pi}(P_{a \mu}^{* \dag} P_b \partial^\mu M_{ba}
+P_a^\dag P_{b \mu}^{*}\partial^\mu M_{ba} \nonumber\\
&+i \epsilon^{\mu\nu\rho\sigma} P_{a \rho}^{* \dag}P_{b
\sigma}^{*}v_\nu \partial_\mu M_{ba}),
\end{eqnarray}
where $O(M^3)$ terms are ignored.

Taking the mass difference between $P_a$ and $P_a^*$ into account,
the following term has to be added into $\cal{L}$ in Eq.(13):
\begin{equation}
  \frac{\lambda_2}{m_Q}{\rm Tr}{\bar H}_a \sigma^{\mu\nu} H_a
  \sigma_{\mu\nu},
\end{equation}
where $\lambda_2$ is a constant containing interaction information
at the quark level.

So the propagators for heavy pseudoscalar and vector mesons are
\begin{equation}
\frac{i}{2(v\cdot p + \frac{3}{4}\triangle)}
\end{equation}
and
\begin{equation}
 \frac{-i(g_{\mu\nu} - v_{\mu\nu})}{2(v\cdot p -
 \frac{1}{4}\triangle)}
\end{equation}
respectively, where $p$ is the residual momentum of the heavy
meson. In Eqs.(17)(18)
\begin{equation}
\triangle = - \frac{8\lambda_2}{m_Q},
\end{equation}
which is the mass difference between vector and pseudoscalar heavy
mesons.
\section{Formulas for the extrapolation of the $B$-meson decay constant}
\label{sec:2}
The decay constant of a  pseudoscalar heavy meson,
$P$, is defined by
\begin{equation}
\langle0|J^\mu|P\rangle = i f_P m_P v^\mu,
\label{jmu}
\end{equation}
where the axial-vector current is
\begin{equation}
J^\mu = \bar{q} \gamma^\mu (1-\gamma_5) Q,
\end{equation}
$q$ denotes $u, d$ light quarks and $Q$ represents heavy quarks
 $c, b$.

 The current in Eq.(21) can be written in the low energy chiral theory as\cite{Mark92}
\begin{equation}
J_a^\mu = (\frac{i\alpha}{2}){\rm Tr}[\gamma^\mu(1 - \gamma_5)H_b
\xi_{ba}^\dag],
\end{equation}
where $\alpha$ is a parameter.

Substituting Eq.(5) into Eq.(22), we have the following explicit
form for the current $J_a^\mu$:
\begin{equation}
J_a^\mu = i \alpha (P_b^{*\mu} - v^\mu P_b) \xi_{ba}^\dag.
\end{equation}

We make a Taylor expansion for $M$ and omit $O(\frac{1}{f_\pi^3})$
terms. This leads to the following expression for $J^\mu_a$:
\begin{equation}
J_a^\mu = i \alpha (P_b^{*\mu} - v^\mu P_b)[\delta_{ba} -
\frac{i}{f_\pi}(M)_{ba} - \frac{1}{2f_\pi^2}(M^2)_{ba}].
\end{equation}

The diagrams for the heavy meson decay constant to one pion loop
order from Eqs.(15)(24) are shown in Fig. 1. We can see that there
are three diagrams for one pion loop corrections to the
pseudoscalar heavy meson decay constant.

From Fig. 1(a), it is easy to obtain
\begin{equation}
\alpha = - f_P^{(0)} \sqrt{m_P} ,
\end{equation}
where $f_P^{(0)}$ is the heavy pseudoscalar meson decay constant
 at the tree level where pion loop corrections are not taken into account.

Fig. 1(b) represents the pion loop correction from the axial
current $J_a^\mu$ itself. It can be expressed as
\begin{equation}
f_P^{(b)} = -\frac{3i}{4f_\pi^2} f_P^{(0)}\int\frac{{\rm d^4
}k}{(2\pi)^4}\frac{1}{k^2-m_\pi^2},
\end{equation}
where $k$ is the four momentum of the pion in the loop and $m_\pi$
is the pion mass which is not necessary to be its physical mass.

Choosing the contour in the lower half complex plane of $k_0$, in
which there is only one pole for $k_0$, we have the following
result after integrating over $k_0$:
\begin{equation}
f_P^{(b)} = -\frac{3}{8f_\pi^2} f_P^{(0)} \int\frac{{\rm d^3
}k}{(2\pi)^3}\frac{1}{w_k},
\end{equation}
where $w_k = \sqrt{|\vec{k}|^2 + m_\pi^2}$ .

Fig. 1(c) vanishes because of the identity $$v^\mu (g_{\mu\nu} -
v_\mu v_\nu) = 0.$$

The contribution of Fig. 1(d) to the matrix element in
Eq.(\ref{jmu}) is
\begin{equation}
\frac{9ig^2}{2(v\cdot p +
\frac{3}{4}\bigtriangleup)f_\pi^2}f_P^{(0)}X
\end{equation}
multiplied by $i m_P v^\mu$, where the self-energy contribution is
proportional to
\begin{equation}
X = \frac{1}{3}\int\frac{{\rm d^4}k}{(2\pi)^4}\frac{k^2 - (v\cdot
k)^2}{[v\cdot(p - k) -\frac{1}{4}\triangle](k^2 - m_\pi^2)}.
\label{x}
\end{equation}

Since $v\cdot p$ is a Lorentz scalar, we are free to choose the
special frame in which the heavy meson is at rest: $v^0 = 1,
\vec{v} = 0$. Choosing the same contour in the evaluation of the
integral as before, we have (details are given in Ref.\cite{guo1})
\begin{equation}
X = \frac{i}{6}\int\frac{{\rm d^3}k}{(2\pi)^3}\frac{|\bf k
|^2}{w_k(w_k-\delta)},
\label{x1}
\end{equation}
where $\delta = v\cdot p - \frac{1}{4}\triangle$.

The contribution to $f_P$ from the one-loop wave function
renormalization, $f_P^{(d)}$, can be obtained from the derivative
of $X$ with respect to $v \cdot p$:
\begin{equation}
   \frac{1}{2} \frac{\partial X}{\partial v\cdot p
  }|_{v\cdot p = - \frac{3}{4}\Delta}.
\end{equation}

Then we have
\begin{equation}
  f_P^{(d)} = - \frac{3g^2}{8f_\pi^2}f_P^{(0)} \int \frac{{\rm
  d^3}k}{(2\pi)^3}\frac{|\bf k|^2}{w_k(w_k-\delta)^2}.
\end{equation}

Adding all the contributions from the diagrams in Fig. 1, we have
the following expression for $f_P$:
\begin{equation}
f_P = f_P^{(0)} + f_P^{(b)} + f_P^{(d)},
\end{equation}
where the first term represents the contribution at the tree
level, and the other two come from one pion loop corrections.

When the light quark mass is near the chiral limit, pion loop
corrections give the dominate contributions to $f_P$; on the
contrary, pion loop contributions vanish in the limit $m_\pi
\rightarrow \infty$. In order to extrapolate the lattice data, we
also have to know the behavior of $f_P$ at the large pion mass.
Based on the non-relativistic constituent quark model, it is known
for a long time that, up to logarithmic corrections,  $f_P$ obeys
the asymptotic scaling law when the constitute quark masses are
very large:
\begin{equation}
 f_P\sqrt{m_P} = {\rm constant}.
\end{equation}
The above behavior  is also obtained in the Dyson-Schwinger
Equation approach\cite{M.A}.

Therefore, it is convenient to extrapolate $f_P\sqrt{m_P}$ instead
of $f_P$. Then we only need to introduce a constant parameter
while fitting the lattice data in the region where $m_\pi$ is
large.

Based on the above argument, we propose the following functional
form for extrapolating the heavy meson decay constant from
unphysical region to the physical limit:
\begin{equation}
  \Phi_P = \sqrt{m_P} (f_P^{(0)} + f_P^{(b)} + f_P^{(d)}) +
  b_P^\prime,
\end{equation}
where
\begin{equation}
\Phi_P = f_P\sqrt{m_P},
\end{equation}
and $ b_P^\prime$ is a parameter.

It is convenient to define the following new parameters:
\begin{eqnarray}
  a_P &=& \sqrt{m_P} f_P^{(0)},\nonumber\\
  b_P &=& \sqrt{m_P} f_P^{(0)} + b_P^\prime.
\end{eqnarray}
So we have the following explicit functional form for
extrapolating the lattice data:
\begin{equation}
\Phi_P = a_P \sigma^\chi_P + b_P,\label{extr}
\end{equation}
where
\begin{eqnarray}
 \sigma^\chi_P &= &-\frac{3}{8f_\pi^2} \int\frac{{\rm
d^3 }k}{(2\pi)^3}\frac{1}{w_k}\nonumber\\
 && - \frac{3g^2}{8f_\pi^2} \int \frac{{\rm
  d^3}k}{(2\pi)^3}\frac{|\bf k|^2}{w_k(w_k-\delta)^2}
\label{sig}
\end{eqnarray}
from Eqs.(27)(32), and $a_P$ and $b_P$ are the parameters to be
determined by fitting the lattice data.

As mentioned in Introduction, for extrapolating the lattice data
which are outside the power-counting regime, the FRR technique can
be used for the resummation of the chiral expansion. In FRR, a
finite-range cutoff  in the momentum integrals of the pion loop
diagrams is introduced. We choose two different approaches for
evaluating the integrals in Eq.(39). One is the sharp-cutoff form,
$\theta (\Lambda-|\vec k|)$, and the other is the dipole form,
$\Lambda^4/(\Lambda^2 + |\vec k|^2)^2$. Both of them make the
integrals converge. When the pion mass is greater than the cutoff
$\Lambda$, which characterizes the finite size of the source of
the pion, the Compton wavelength of the pion is smaller than that
of the source and pion loop contributions are suppressed as powers
of $\Lambda/m_\pi$. Obviously the dipole form is more realistic.
Since the leading nonanalytic contribution of pion loops is only
associated with the infrared behavior of the integrals in Eq.(39),
it does not depend on the details of $\Lambda$.

Since the lattice simulations are performed on a finite volume
grid, the finite size effects should be taken into
account\cite{D.B}. In the finite periodic volume, the available
momenta $k$ are discrete:
\begin{equation}
  k = \frac{2\pi n}{aL}, \label{k}
\end{equation}
where $L$ is the number of lattice sites in the $x, y, z$
direction, and the integer $n$ is in the following range:
\begin{equation}
  -\frac{L}{2} < n \leq \frac{L}{2}.
\end{equation}
Since the pion momenta on the finite lattice volume are discrete,
we should take this into account by replacing the continuous
integral over $k$ in Eq.(\ref{sig}) with a discrete sum over $k$,
\begin{equation}
  \int{\rm d^3}k\approx (\frac{2\pi}{aL})^3
  \sum_{k_x,k_y,k_z}\ ,
\end{equation}
where the discrete momenta $k_x$, $k_y$, $k_z$ are given in
Eq.(40).  From Ref.\cite{Alan1}, the volume of lattice simulation
is $20^3 \times 64$, corresponding to $L = 20$. The smallest
momentum allowed on the lattice is $2\pi / aL$.
\section{Extrapolation of lattice data for pseudoscalar heavy meson decay constant}
\label{sec:4}

Lattice gauge theory is the only quantitative tool currently
available to calculate nonperturbative phenomena in QCD from first
principles. Quark vacuum polarization is the most expensive
ingredient in lattice QCD simulations, especially when quark
masses are very small, as in the case of $u$ and $d$ quarks. The
MILC Collaboration has established unquenched gauge configuration,
which contains three flavors of light sea quarks to include the
effects of realistic quark vacuum polarization\cite{C.B}. This
staggered quark discretization of QCD has several advantages,
which are offset by the fact that staggered quarks always come in
groups of four identical flavors. In
Refs.\cite{S.N}\cite{C.A}\cite{C.A1}, the Kogut-Susskind action
with improved flavor and rotational symmetry suitable for
dynamical fermion simulations is constructed. At tree level, the
action has no couplings of quarks to gluons with a transverse
momentum component $\pi / a$. So the flavor symmetry violating
terms in the action are completely removed at tree level. Then the
rotational symmetry is improved by introducing the Naik term.
Finally, an extra five link staple is introduced in order to
cancel errors of ${\cal O}(a^2 g^2)$ from the fattening. The
result action is called $Asq$ action. It can be further improved
by tadpole improvement, leading to $Asqtad$, which is an order
${\cal O} (a^4, a^2 g^2)$ accurate fermion action.

The authors of Ref.\cite{Alan} studied the $B$-meson decay
constant employing the MILC Collaboration unquenched gauge
configuration, and light sea quarks are represented by the highly
improved staggered quark action. The good chiral properties of the
latter action allow for a much smoother chiral extrapolation to
the physical region. The $b$ quark inside $B$ meson is treated
only as the valence quark in simulations because its effect in the
sea should be suppressed by inverse powers of the $b$ quark
mass\cite{M.N}. NRQCD lattice action was used to deal with the
valence $b$ quark, which has been developed over many years
\cite{B.A}\cite{G.P}\cite{C.T}. The $b$ quark is non-relativistic
inside its bound state, hence a non-relativistic expansion of the
QCD action is appropriate which accurately handles scales of the
order of typical momenta and kinetic energies inside these states.

Some lattice data of $\Phi_B$ are obtained in Ref.\cite{Alan} from
full or partially quenched QCD. The coarse lattice spacing $a$ is
around 0.12fm. In the case of full QCD, dynamical light quark
masses are chosen as $m_q /m_s$ = 0.125, 0.175, 0.25 and 0.5,
where $m_s$ is the strange quark mass, corresponding to $\Phi_B$ =
0.516(5)(15), 0.519(5)(15), 0.517(8)(15) and 0.540(5)(15) (in unit
of GeV$^{3/2}$), respectively (the errors in the first parentheses
are statistical and the second come from lattice spacing
uncertainties). In the case of partially quenched QCD, dynamical
light quark masses are chosen as $m_q /m_s$ = 0.125 and 0.5,
corresponding to $\Phi_B$ = 0.506(5)(14), and 0.547(5)(15) (in
unit of GeV$^{3/2}$), respectively. Two values of $\Phi_B$ in the
case of fine lattices with $a$ being around 0.087fm have also been
accumulated, where the staggered valence light propagators created
by the Fermilab Collaboration were used to treat the light quarks.

We choose to extrapolate the coarse data (full and partially
quenched) since the number of the fine data is only two. If we
used the fine data, the uncertainty of the fitted results due to
the error in the lattice data would be too large. Explicit lattice
simulations show that over the rang of mass of interest to us,
where the pion mass is not constrained by the chiral limit,
$m_\pi$ is proportional to $m_q$\cite{C.A2}.

In our model, there are two parameters, $a_P$ and $b_P$, to be
fixed in Eq.(\ref{extr}). These parameters are related to $g$,
$\lambda_2$ and $\Lambda$. $g$ and $\lambda_2$ represent the
interaction at the quark level and cannot be determined from the
chiral perturbation theory for heavy mesons. From the experimental
data for the decay width for $D \rightarrow D^*\pi$, we have the
variation of $g$ between 0.51 and 0.67\cite{A.Ana}. From Eq.(16),
$\lambda_2$ is related to the mass splitting between a heavy
vector meson and a heavy pseudoscalar meson. Based on the
experimental data for $B$ mesons, the value of $\lambda_2$ should
be around $-0.03 GeV^2$. To see the dependence on $\lambda_2$ in
our fits, we let $\lambda_2$ vary between $-0.03 GeV^2$ and $-0.02
GeV^2$. We let $\Lambda$ vary between $0.4 GeV$ and $0.6 GeV$,
considering the scale of the finite size of the pion
source\cite{guo1}\cite{guo2}. The two parameters, $a_P$ and $b_P$,
are determined by the least squares fitting method in our fit.

We first extrapolate the full QCD data. In this case, the
uncertainties of the parameters $a_B$ and $b_B$ are around 90\%
and 10\%, respectively, due to the errors in the lattice data. The
large uncertainty of $a_B$, associated with the chiral correction
term, is due to the lack of the lattice data near the physical
region. In Table 1, the fitted results in Columns 3 and 5 (Columns
2 and 4) correspond to the case  where the finite lattice volume
effects are (not) taken into account. We can see that the
difference between these two cases is not very large.

The fitted result for $f_B$ is obtained from Column 5 in Table 1,
where the dipole form factor is used and the correction from the
discretization of the pion momentum  is considered:
\begin{equation}
  f_B = 214.2\pm 13.3 \pm 0.8  MeV,
\label{fb}
\end{equation}
where the first uncertainty is from the errors in the lattice
data, and the second is from the variation of parameters ($g,
\lambda_2, \Lambda$) in our model.

If we use the sharp cutoff scheme, the fitted result for $f_B$ is
\begin{equation}
  f_B = 215.1\pm 15.8 \pm 4.1 MeV.
\end{equation}

We then extrapolate the full QCD data together with the two
partially quenched data (although, in principle, one should use
quenched chiral Lagrangian to extrapolate these two data). As
expected, we find that the uncertainties which are due to the
errors in the lattice data for $a_B$, $b_B$, and the extrapolated
$B$-meson decay constant are all reduced since the number of the
lattice data increases (see Table 2).

The fitted result for $f_B$ is obtained from Column 5 in Table 2,
where the dipole form factor is used and the correction from the
discretization of the pion momentum  is considered as before:
\begin{equation}
  f_B = 209.4\pm 9.7 \pm 1.0  MeV,
\label{ff0}
\end{equation} where the first uncertainty is from the
errors in the lattice data, and the second is from the variation
of parameters ($g, \lambda_2, \Lambda$) in our model.

If we use the sharp cutoff scheme, the fitted result for $f_B$ is
\begin{equation}
  f_B = 210.5\pm 11.4 \pm 5.7 MeV.
\label{ff}
\end{equation}

It can be seen that the extrapolated results for the $B$-meson
decay constant are consistent with each other in the above two
extrapolation cases and in the two cutoff schemes if we consider
the uncertainties of the extrapolated results. We can also see
that the uncertainties due to variations of our model parameters
in the dipole scheme, which is more realistic, are smaller than
those in the sharp cutoff scheme.

\section{Summary and discussion}
\label{sec:5} It is known that the strong interactions are
constrained by the heavy quark symmetry when the heavy quark mass
goes to infinity, and by the chiral symmetry when the light quark
mass approaches zero. The chiral perturbation theory can be
applied when the light quark mass (or the pion mass) is small
enough. In the region where the light quark mass is bigger than
about 60 MeV, the non-relativistic constituent quark model is
appropriate. Base on these, we propose a phenomenological
functional form to extrapolate the lattice data for the $B$-meson
decay constant to the physical region, combining the chiral
perturbation theory and the non-relativistic constituent quark
model.

We evaluate pion loop contributions when $m_\pi$ is small with the
aid the chiral perturbation theory for heavy mesons. This leads to
correct nonanalytic chiral behavior of $f_B$ in the chiral limit.
Since the lattice data for the $B$-meson decay constant is outside
the power-counting regime, instead of simply using the chiral
logarithms as done in Ref.\cite{Alan}, we employ the finite-range
regularization technique in order to resum higher order terms of
the chiral expansion. At large pion masses, we introduce a
constant parameter to fit the lattice data for $\Phi_B$ (which
appear in the region where the constituent quark model is suitable
to be applied) based on the non-relativistic constituent quark
model. This is also different from the term linear in light quark
mass used in Ref.\cite{Alan}. It has been pointed out that in
large pion mass region, the extrapolation based on the constituent
quark model is more reasonable than the simple linear
extrapolation\cite{guo2}. The finite lattice volume effects are
taken into account by replacing the continuous integral with a
discrete sum. In our fit, the two parameters are determined by the
least squares fitting method.

In Ref.\cite{Alan}, the lattice data for the $B$-meson decay
constant (both full QCD and partially quenched data) are
extrapolated based on the staggered chiral perturbation theory
with power counting being used. The result of the extrapolated $B$
decay constant at the physical pion mass is
\begin{equation}
 f_B = 0.216 (9) (19) (4) (6) GeV,
\label{ffa}
\end{equation}
where the errors, from left to right,
are statistics plus scale plus chiral extrapolations, higher order
matching, discretization, and relativistic corrections plus $m_b$
tuning respectively.

It can be seen that the central values for the extrapolated
$B$-meson decay constant in Eqs.(\ref{ff0})(\ref{ff}) are smaller
than that in Eq.(\ref{ffa}).  However, because of the large
uncertainties due to the errors in the lattice data and the model
parameters etc. the extrapolated $B$-meson decay constant in our
work is still consistent with that in Ref.\cite{Alan}.

Now we discuss the uncertainties in our  model. We have three
parameters, $\lambda_2$, $g$ and $\Lambda$, where the first two
are related to the color-magnetic-moment operator at order $1/m_Q$
in HQET and the interaction between  heavy mesons and Goldstones
in chiral perturbation theory, respectively. Appropriate ranges
for them (obtained from the comparison with the experimental data)
are used in our fit process. We did not take into account the
$O(a^2)$ contributions in the Kogut-Susskind action, which could
give 7\% corrections to our result\cite{Alan}. In our approach,
sea quark mass is not extrapolated. Usually quenched lattice QCD
gives about 90\% contribution to physical quantities, hence sea
quark contribution is at about 10\% level. Therefore, one expects
the uncertainty from sea quark extrapolation is about $1 \sim
2$\%. Taking these into account, our extrapolated result in
Eqs.(\ref{fb}-\ref{ff}) may have another 8\% uncertainty. In our
future work we will investigate these issues carefully in order to
reduce these uncertainties.

{\it Acknowledgements.} This work was supported in part by Special
Grants for "Jing Shi Scholar" of Beijing Normal University and by
National Nature Science Foundation of China Project No. 10675022.

\onecolumn
\begin{table}[htb]
\begin{center}
\caption{Fitting parameters, the fitted results in the
sharp-cutoff and the dipole schemes and the (relative)
uncertainties caused by the errors in the lattice data in the
 full QCD case are shown.
The variations of the values in the second to the fifth columns
are due to the variations of the parameters in our model.}
\begin{tabular}{lllll}
\hline\hline
Form factor &sharp cutoff &{}&dipole &{}  \\
 \hline
Volume & infinite &finite & infinite & finite \\
\hline
$a_B$ (GeV$^{3/2}$) &0.3644-0.9395 &0.3759-1.0388 &0.5068-1.1943 &0.5098-1.1965\\
 $\triangle a_B / a_B$ (\%) &89.30-91.38 &89.36-91.62 &88.72-89.83 &88.75-89.84 \\
 $b_B$ (GeV$^{3/2}$) &0.5715-0.5928 &0.5712-0.5910 &0.6017-0.6368 &0.5981-0.6275\\
 $\triangle b_B / b_B$ (\%) &7.86-10.58 &7.69-1.05 &11.80-15.88 &11.34-14.81\\
$\Phi$ (GeV$^{3/2}$) &0.4944-0.4994 &0.4844-0.5035 &0.4971-0.5000 &0.4900-0.4936\\
 $\triangle \Phi$ (GeV$^{3/2}$) &0.0227-0.0275 &0.0195-0.0364 &0.0221-0.0248 &0.0277-0.0305\\
 $\triangle \Phi / \Phi$ (\%)&4.54-5.57 &3.87-7.52 &4.42-4.99 &5.61-6.23\\
$f_B$ (GeV) &0.2153-0.2175 &0.2110-0.2193 &0.2165-0.2178 &0.2134-0.215\\
 $\triangle f_B$ (GeV) &0.0099-0.0120 &0.0085-0.0159 &0.0096-0.0108 &0.0120-0.0133\\
 $ \triangle f_B  / f_B$ (\%)&4.54-5.57 &3.87-7.52 &4.42-4.99 &5.61-6.23\\
\hline\hline
\end{tabular}
\end{center}
\end{table}

\begin{table}[htb]
\caption{Fitting parameters, the fitted results in the
sharp-cutoff and the dipole schemes and the (relative)
uncertainties caused by the errors in the lattice data in the case
where both the full QCD and partially quenched QCD data are fitted
are shown. The variations of the values in the second to the fifth
columns are due to variation of the parameters in our model. }
\begin{center}
\begin{tabular}{lcccc}
\hline \hline
Form factor &sharp cutoff &sharp cutoff&dipole &dipole  \\
 \hline
Volume & infinite &finite & infinite & finite \\
\hline
$a_B$ (GeV$^{3/2}$) &0.5160-1.3369 &0.5075-1.4148 &0.6810-1.6116 &0.6851-1.6146\\
 $\triangle a_B / a_B$ (\%) &49.47-52.91 &47.35-47.78 &47.23-47.43 &47.24-47.44 \\
 $b_B$ (GeV$^{3/2}$) &0.5944-0.6238 &0.5878-0.6160 &0.6295-0.6758 &0.6247-0.6633\\
 $\triangle b_B / b_B$ (\%) &6.22-8.69 &5.31-7.16 &8.03-10.66 &7.72-9.98\\
 $\Phi_B$ (GeV$^{3/2}$) &0.4839-0.4910 &0.4707-0.4968 &0.4971-0.5000 &0.4788-0.4835\\
 $\triangle \Phi_B$ (GeV$^{3/2}$) &0.0177-0.0217 &0.0144-0.0262 &0.0163-0.0182 &0.0202-0.0223\\
 $\triangle \Phi_B / \Phi_B$ (\%)&3.62-4.48 &2.90-5.58 &3.32-3.72 &4.18-4.66\\
 $f_B$ (GeV) &0.2105-0.2137 &0.2048-0.2162 &0.2163-0.2176 &0.2084-0.2104\\
 $\triangle f_B$ (GeV) &0.0077-0.0094 &0.0063-0.0114 &0.0071-0.0079 &0.0088-0.0097\\
 $\triangle f_B / f_B$ (\%)&3.62-4.48 &2.90-5.58 &3.32-3.72 &4.18-4.66\\
\hline \hline
\end{tabular}
\end{center}
\end{table}

\begin{figure}
\begin{center}
\caption{Heavy meson decay constant to one pion loop order. The
black squares represent the weak current in Eq.(22), and dashed
lines denote the pion.}
\resizebox{0.5\textwidth}{!}{%
  \includegraphics{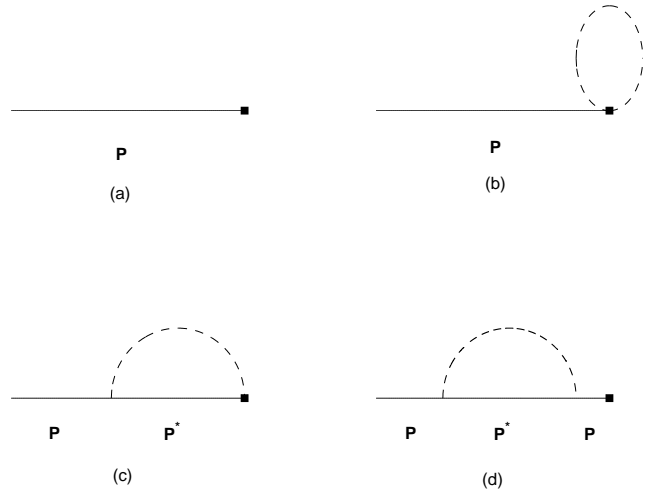}
}
\label{fig:1}       
\end{center}
\end{figure}
\begin{figure}
\caption{Extrapolation of the full QCD lattice data for the
$B$-meson decay constant. Dashed line (dotted line) corresponds to
the sharp cutoff scheme and infinite (finite) volume; dash-dot
line (solid line) corresponds to the dipole scheme and infinite
(finite) volume. The vertical solid line corresponds to the
physical pion mass.}
\resizebox{1.0\textwidth}{!}{%
  \includegraphics{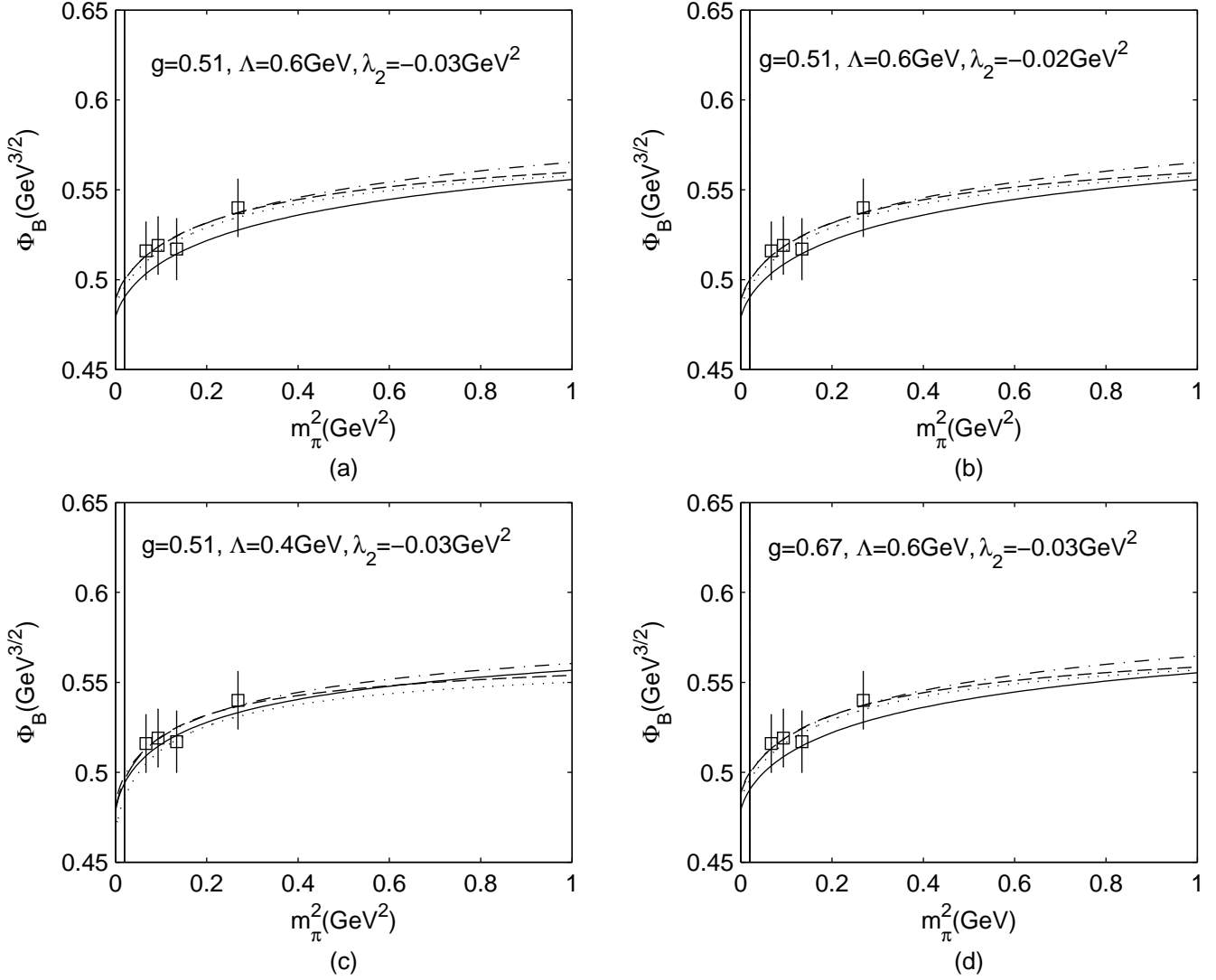}
}
\label{fig:2}       
\end{figure}

\begin{figure}
\caption{Extrapolation of both the full QCD and partially quenched
QCD lattice data (which are represented by squares and diamonds
respectively) for the $B$-meson decay constant. Dashed line
(dotted line) corresponds to the sharp cutoff scheme and infinite
(finite) volume; dash-dot line (solid line) corresponds to the
dipole scheme and infinite (finite) volume. The vertical solid
line corresponds to the physical pion mass.}
\resizebox{1.0\textwidth}{!}{%
  \includegraphics{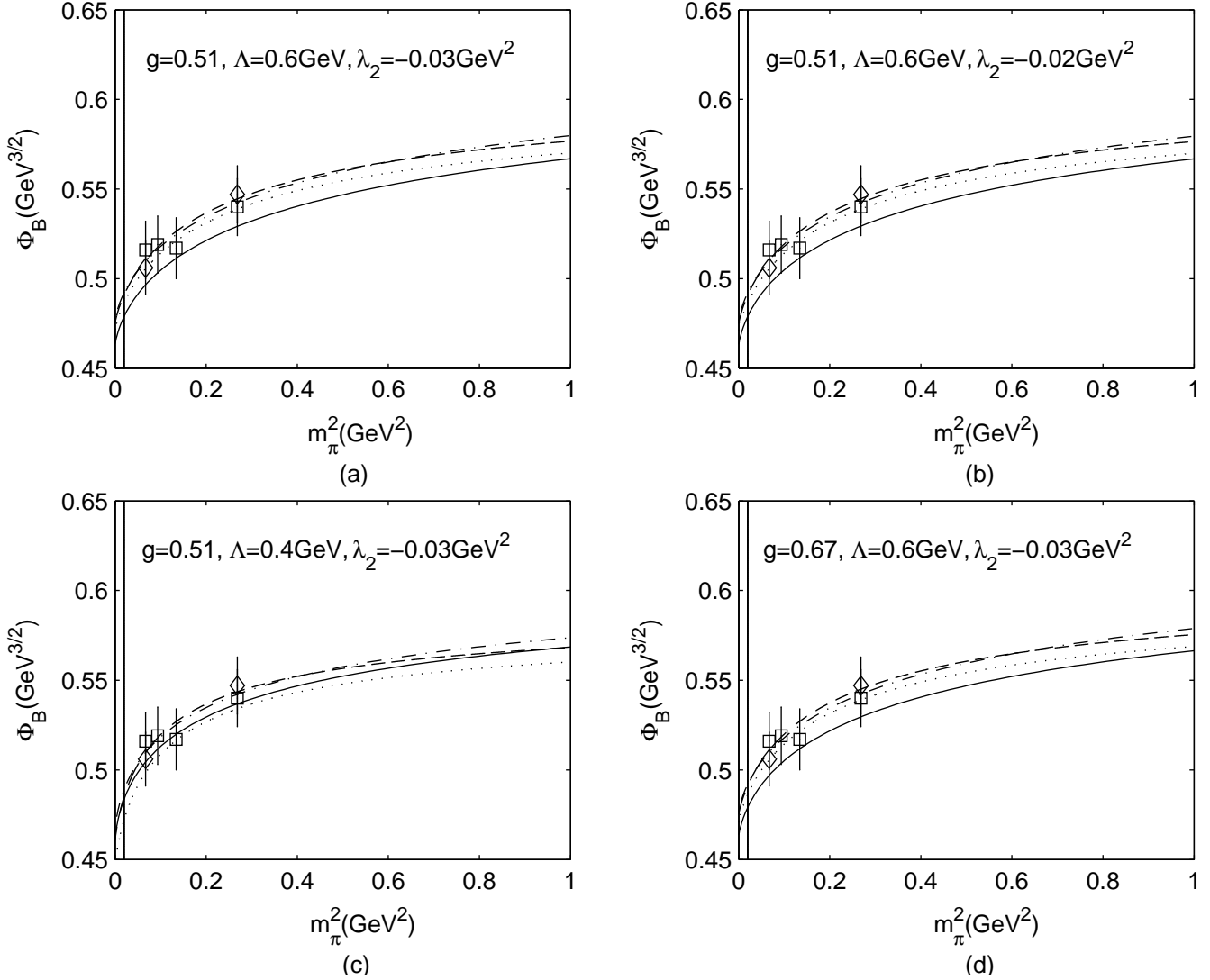}
}
\label{fig:3}       
\end{figure}

\end{document}